\documentstyle[11pt,epsf]{article}
\newcommand{\half}{\mbox{\small{$\frac{1}{2}$}}} 
  
\newcommand{\la}{\langle}
\newcommand{\ra}{\rangle}   
\newcommand{\MSbar}{\overline{\mbox{MS}}} 
\begin{document}
\title{Non-zeta knots in the renormalization of the Wess-Zumino model?} 
\author{P.M. Ferreira \& J.A. Gracey, \\ Theoretical Physics Division, \\ 
Department of Mathematical Sciences, \\ University of Liverpool, \\ 
Peach Street, \\ Liverpool, \\ L69 7ZF, \\ United Kingdom.}
\date{} 
\maketitle
\vspace{5cm}
\noindent
{\bf Abstract.} We solve the Schwinger Dyson equations of the $O(N)$ symmetric
Wess-Zumino model at $O(1/N^3)$ at the non-trivial fixed point of the 
$d$-dimensional $\beta$-function and deduce a critical exponent for the wave
function renormalization at this order. By developing the $\epsilon$-expansion 
of the result, which agrees with known perturbation theory, we examine the 
distribution of transcendental coefficients and show that only the Riemann
$\zeta$ series arises at this order in $1/N$. Unlike the analogous calculation 
at the same order in the bosonic $O(N)$ $\phi^4$-theory non-zeta 
transcendentals, associated with for example the $(3,4)$-torus knot, cancel.  

\vspace{-19cm} 
\hspace{13.5cm} 
{\bf LTH-418} 

\newpage

One of the more exciting developments in renormalization theory in recent years
has been the realization that there is a relation between the Riemann zeta and 
non-zeta transcendental numbers which appear in the renormalization group 
functions of a variety of theories and the positive knots of mathematics,
\cite{dk1,dk2,dkdb}. Briefly, in \cite{dk1} it was demonstrated that a 
multiloop Feynman diagram could be easily associated with a link diagram which,
after applying basic knot manipulations and a skeining relation, could be 
reduced to an underlying positive knot listed in the classification tables, 
plus trivial unknots. It turned out that one could build up a one-to-one 
correspondence between different positive knots which would appear in a Feynman
diagram and the irrational numbers, like the Riemann zeta series, $\zeta(n)$, 
which appeared in the renormalization constant of the {\em same} graph. For 
example, if a diagram contained the number $\zeta(3)$ in its simple pole with 
respect to the regularization then it always occurred in a Feynman diagram 
whose associated link diagram reduced to the trefoil or $(2,3)$-torus knot. 
More generally, the presence of $\zeta(2n-3)$, for $n$ $\geq$ $3$, in a 
renormalization constant corresponds to a $(2,2n-3)$-torus knot in the link 
diagram of the original Feynman diagram, \cite{dk1,dk2}. Indeed intense 
investigation revealed that other {\em non}-zeta transcendentals which are 
represented by independent double and triple infinite series were $1$-$1$ 
associated with other sets of torus knots, \cite{dkdb}. Moreover it was found 
that there was an intimate relation between the braid word representation of 
the knot and the structure of its knot number when it is expressed as an 
infinite sum. This empirical evidence has been examined with great intensity 
more recently and has so far remained robust, \cite{algnum}. An overview of the
status of the area can be found in \cite{dkbook}. One advantage for 
renormalization theory of having such an association is that one immediately 
knows from the simple link diagram of a Feynman diagram which independent 
transcendentals to expect in its calculation. This can therefore be adapted as 
a calculational aid for the very high order loop diagrams in which they arise. 
For example, performing an algebraic-numerical calculation one can project onto
the appropriate number basis for the diagram. Indeed this method was 
successfully used in \cite{dkdb} to compute all the primitively divergent 
graphs at six and seven loop which contribute to the $\MSbar$ $\beta$-function 
of $\phi^4$ theory. This was the first order in this theory to involve these 
new non-zeta transcendentals. 

Clearly to gain some insight into the underlying knot theory requires very
high order multiloop calculations which is a technical and difficult exercise, 
\cite{dk1,dkdb,algnum}. However, in \cite{bgk} non-zeta transcendentals which 
occur in graphs with subgraph divergences were accessed in the wave function 
renormalization of $O(N)$ $\phi^4$ theory through the large $N$ expansion. (The
knot relation in graphs with subgraph divergences has also been investigated in
perturbation theory in \cite{dksub}.) A result had been available in 
\cite{vas3} in $d$-dimensions at $O(1/N^3)$ for the critical exponent $\eta$ 
which is related to the wave function anomalous dimension through the critical 
renormalization group. It contained the value of a two loop integral which when
expanded in powers of $\epsilon$, where $d$ $=$ $4$ $-$ $2\epsilon$, contained 
non-zeta transcendentals. The first of these had been known for a long time and
had been denoted in \cite{djbu62a} by $U_{62}$ $=$ $\sum_{n>m>0}^\infty \, 
(-1)^{n-m}n^{-6}m^{-2}$. However, it is now associated in the more modern 
approach with the $(3,4)$-torus knot, \cite{dkdb}, and hence through the braid 
word representation of this knot $U_{62}$ has been replaced by the independent,
but more systematically defined, non-alternating double sum $F_{53}$ $=$ 
$\sum_{n>m>0}^\infty \, n^{-5} m^{-3}$. This discovery yielded the expansion of
the two loop integral to much higher order, \cite{bgk}, and established the 
knot association to a new number of loops. It also provided a basis for making 
a stronger connection with algebraic number theory, \cite{algnum}. Although 
most of the results we have mentioned have been established in simple scalar 
theories the knot approach has proved useful in gaining insight into long term 
multiloop problems in theories with symmetries. Indeed the cancellation of 
$\zeta(3)$ in the quenched $\beta$-function of QED at three loops has been 
explained and extended to higher order cancellations by knot theoretic 
arguments in \cite{quen}. Therefore it would seem that the presence of a 
symmetry in a theory, such as gauge symmetry, can in some way suppress the 
appearance of these transcendental numbers. Moreover it would be interesting to
understand what effect other symmetries, such as {\em four} dimensional 
supersymmetry, will have on the structure of the renormalization group 
functions. This is the subject of this paper where we will study the 
Wess-Zumino model, \cite{wz}, with an $O(N)$ symmetry to $O(1/N^3)$ in a large 
$N$ expansion. The aim is to first calculate the critical exponent, $\eta$, 
corresponding to the wave function renormalization and then to examine the 
expression obtained for it to ascertain how supersymmetry modifies the 
appearance of zeta and non-zeta transcendentals compared with the underlying 
bosonic $\phi^4$-theory. This would therefore extend some of the analysis of
\cite{bgk}. We note that in the underlying bosonic model, $O(N)$ $\phi^4$ 
theory, it was possible to establish not only the first location of $U_{62}$ in
the wave function renormalization at $O(1/N^3)$ but also to provide its 
explicit coefficient, \cite{bgk,nim}. Knowledge of this is clearly important if
one is ever to have a systematic knot based method to perform the 
renormalization since the coefficients of the transcendentals that appear in 
some renormalization scheme will also need to be determined. Therefore the 
large $N$ results will provide important, non-trivial and independent checks. 
We recall that large $N$ calculations in models such as the Wess-Zumino model 
are possible at $O(1/N^2)$ through critical renormalization group methods and 
solution of the Schwinger Dyson equations order by order in $1/N$ at a 
non-trivial fixed point of the $d$-dimensional $\beta$-function. The method had
been originally applied to the $O(N)$ bosonic $\sigma$ model in an impressive 
series of articles, \cite{vas12,vas3}. More recently, similar calculations in 
supersymmetric theories have been made possible in \cite{pfjag} by extending 
the methods of \cite{vas12} to superspace for the $O(N)$ Wess-Zumino model 
which, it is hoped, can be extended to study supersymmetric gauge theories. 

We now turn to the specifics of calculating $\eta$ in the Wess-Zumino model 
with an $O(N)$ symmetry and note that its action in superspace is given by 
\begin{equation}
S ~=~ \int \, d^d x \, \left[ \, \int d^4 \theta \, \left( 
\bar{\Phi}^i \Phi^i + \frac{\bar{\sigma}\sigma}{g^2} \right) ~-~ \frac{1}{2}
\int d^2 \theta \, \sigma{\Phi}^2 ~-~ \frac{1}{2} \int d^2 \bar{\theta} \, 
\bar{\sigma} \bar{\Phi}^2 \right]  
\end{equation}
where $\Phi^i$ and $\sigma$ are chiral superfields, $1$ $\leq$ $i$ $\leq$ $N$
and $g$ is the coupling constant which has been rescaled into the $\sigma$ 
kinetic term to ensure that the interaction in $d$-dimensions is in a form for
which the conformal integration technique or uniqueness, \cite{vas12}, can be 
readily applied to ease the calculation of high order integrals. The four loop 
$d$-dimensional $\beta$-function of the model is deduced from the result for a
general symmetry group given in \cite{wzgen} which used the non-renormalization
theorem, \cite{wz,nonren}, and generalized earlier results, 
\cite{wz3loop,avdeev}. Specifying to an $O(N)$ group it is given by, 
\cite{wzgen},  
\begin{eqnarray}
\beta(g) &=& \frac{1}{2} (d-4)g ~+~ \frac{1}{2} (N+4) g^2 ~-~ 2 (N+1)g^3 
{}~+\, \left( \frac{1}{2} (N^2+11N+4) + 3(N+4)\zeta(3) \right) g^4 \nonumber \\ 
&& +~ \left( \frac{1}{6} (N^3 - 36N^2 - 84N - 20) - 3(N^2+16N+24)\zeta(3) 
\right. \nonumber \\ 
&& \left. ~~~~~~ +~ \frac{3}{2}(N+4)^2\zeta(4) - 10(N+2)(N+4)\zeta(5) 
\frac{}{} \right) g^5 ~+~ O(g^6) 
\label{beta} 
\end{eqnarray}
where $\zeta(n)$ is the Riemann zeta function. As we will be calculating 
information relevant to the $\Phi$-superfield renormalization, we note also 
that, \cite{wzgen},  
\begin{eqnarray}
\gamma_{\Phi}(g) &=& g ~-~ \frac{1}{2} (N+2) g^2 ~-~ \frac{1}{4} 
\left( N^2 - 10N - 4 - 24\zeta(3) \right) g^3 \nonumber \\  
&&-~ \left( \frac{1}{24} (3N^3 + 16N^2 + 152N + 40) - \frac{\zeta(3)}{4} 
(N^2 - 4N - 36)(N+4) \right. \nonumber \\ 
&& \left. ~~~~~~ -~ 3\zeta(4)(N+4) + 20\zeta(5)(N+2) \frac{}{} \right) g^4
{} ~+~ O(g^5)  
\label{gamphi}
\end{eqnarray}
The $d$-dimensional fixed point where we will analyse the Schwinger Dyson 
equation is given by the non-trivial zero of the $\beta$-function, 
(\ref{beta}). Specifically expanding in powers of $1/N$ it is given by,  
\cite{pfjag}, 
\begin{eqnarray} 
g_c &=& \frac{2\epsilon}{N} ~+~ \left( - \, 8\epsilon + 16\epsilon^2 
- 8\epsilon^3 - \frac{16}{3}\epsilon^4 + O(\epsilon^5) \right) \frac{1}{N^2} 
\nonumber \\  
&& +~ \left( \frac{}{} 32\epsilon - 176\epsilon^2 - 8(6\zeta(3) - 37)\epsilon^3
\right. \nonumber \\ 
&& \left. ~~~~~~ + \frac{16}{3}( 60\zeta(5) - 9\zeta(4) + 18\zeta(3) - 4 ) 
\epsilon^4 + O(\epsilon^5) \right) \frac{1}{N^3} ~+~ 
O \left( \frac{\epsilon}{N^4} \right) 
\end{eqnarray} 
Evaluating the renormalization group functions at $g_c$ defines the critical 
exponents. So, for example, $\eta$ $=$ $\gamma_\Phi(g_c)/2$. In \cite{pfjag} an 
expression for $\eta$ was deduced at $O(1/N^2)$ by replacing the lines of the
Schwinger Dyson equation, truncated at the same order, by the asymptotic
scaling forms near $g_c$ of the propagators of the respective fields. For an  
Euclidean space-time these are, 
\begin{eqnarray}
\la \bar{\Phi}(-p,\theta) \Phi(p,\theta^\prime) \ra & \sim & 
\frac{A \delta^4 (\theta - \theta^\prime)}{(p^2)^{\mu - \alpha}} \nonumber \\  
\la \bar{\sigma}(-p,\theta) \sigma(p,\theta^\prime) \ra & \sim & 
\frac{B \delta^4 (\theta - \theta^\prime)}{(p^2)^{\mu - \beta}} 
\end{eqnarray}
as $p$ $\rightarrow$ $\infty$, where we have set $d$ $=$ $2\mu$, $A$ and $B$ 
are unknown momentum independent amplitudes and  
\begin{equation}
\alpha ~=~ \mu ~-~ 1 ~+~ \half \eta ~~~,~~~ \beta ~=~ 1 ~-~ \eta 
\end{equation} 
are the exponents of the $\Phi^i$ and $\sigma$ superfields. Their canonical 
dimensions are fixed by ensuring that each term in the $d$-dimensional
superspace action is consistent with $S$ being dimensionless. Due to the
non-renormalization theorem, \cite{wz,nonren}, there is no vertex 
renormalization and its associated exponent, ordinarily denoted by $\chi$, is 
zero to all orders in $1/N$. In particular in the critical region the $\Phi$ 
equation of fig. 1 becomes 
\begin{equation}
0 ~=~ a(\alpha-\mu+1) ~+~ a(\alpha)z  
\end{equation} 
where $z$ $=$ $a^2(\mu-\alpha)a(\mu-\beta)A^2B$ and $a(x)$ $=$ 
$\Gamma(\mu-x)/\Gamma(x)$. The $\sigma$ equation is 
\begin{equation} 
0 ~=~ a(\beta-\mu+1) ~+~ \half N z a(\beta) 
\end{equation} 
where we have chosen to express each representation of fig. 1 in coordinate 
space through a Fourier transform. Thus eliminating $z$ gives 
\begin{equation}
\eta ~=~ \frac{4a(\alpha)a(\beta-\mu+1)}{N(2\mu-2-\alpha)a(\alpha-\mu+2) 
a(\beta)}  
\label{etait} 
\end{equation}
Due to the chirality property of the superfields no new two and three loop
graphs occur at $O(1/N^2)$ unlike the bosonic model and therefore one merely 
expands out (\ref{etait}) to $O(1/N^2)$ to deduce $\eta_2$, where  
$\eta$ $=$ $\sum_{i=1}^{\infty} \eta_i/N^i$. Thus, \cite{pfjag},  
\begin{eqnarray}
\eta_1 &=& \frac{4\Gamma(2\mu-2)}{\Gamma(\mu) \Gamma^2(\mu-1) \Gamma(2-\mu)}
\nonumber \\ 
\eta_2 &=& \left[ \psi(2-\mu) ~+~ \psi(2\mu-2) ~-~ \psi(\mu-1) ~-~ \psi(1) ~+~ 
\frac{1}{2(\mu-1)}\right] \eta_1^2 
\label{eta} 
\end{eqnarray}
where $\psi(x)$ is the logarithmic derivative of the Euler $\Gamma$-function. 

The absence of these higher order graphs and the finiteness of the vertex means
that one can use the Schwinger Dyson approach to compute $\eta_3$ which we 
believe is the first instance this has been attempted. Ordinarily one has to 
use the conformal bootstrap method, \cite{confboot,vas3}, to do this. For the 
Wess-Zumino model, however, it does not seem possible to apply it since that
method relies on the vertex dimension being non-zero. Moreover the chirality 
of the fields ensures that there are only two new topologies at $O(1/N^3)$  
and these are given for the respective Schwinger Dyson equations in figs 2 
and 3. If we denote the values of the graphs for the $\Phi$ equation by 
$\Sigma_i$ and those for the $\sigma$ equation by $\Pi_i$ then the respective
Dyson equations to $O(1/N^3)$ are now 
\begin{eqnarray} 
0 &=& \frac{a(\alpha-\mu+1)}{a(\alpha)} ~+~ z ~+~ z^3 a^5(\alpha)a^3(\beta) 
a(\mu-\alpha-\beta) \Sigma_1 ~+~ N z^4 a^7(\alpha) a^4(\beta) 
a(\mu-\alpha-\beta) \Sigma_2 \nonumber \\ 
0 &=& \frac{2a(\beta-\mu+1)}{Na(\beta)} ~+~ z ~+~ z^3 a^6(\alpha) a^2(\beta) 
a(\mu-2\alpha) \Pi_1 ~+~ N z^4 a^8(\alpha) a^3(\beta) a(\mu-2\alpha) \Pi_2 
\label{eta3sd} 
\end{eqnarray}  
We note that although the factor $a(\mu-\alpha-\beta)$ that appears with each
of the higher order corrections in the $\Phi$ Dyson equation is singular 
when expanded in powers of $1/N$ it will turn out that in the computation of
the diagrams in momentum space there will be a compensating factor of 
$a(\alpha+\beta)$ in its final value. Therefore the final overall contribution  
will be free of infinities. 

The finiteness of the vertex means that one does not need to perform the usual 
renormalization of the Schwinger Dyson equations. Therefore it only remains
to compute the four graphs. As is usual in superfield calculations the first
step in such an evaluation is to carry out the manipulation of the 
supercovariant derivatives present in the super-Feynman rules which is known as
the $D$-algebra. (For an introduction see, for example, \cite{super}.) For the 
topologies illustrated in figs 2 and 3 this will result in the distribution of 
$\Box$-operators in momentum space on various lines. For the non-planar 
topology the two upper lines joining to the external vertex are each modified 
by a $\Box$-operator. For the four loop topology the corresponding lines also 
gain the same operator as well as the completely internal horizontal 
propagator. The effect of a $\Box$-operator is to reduce the power of the 
exponent of that line by unity. For a $\phi$ line this means the power of its 
propagator will vanish whereas that of a $\sigma$ line will be non-zero. One 
might expect that for the graphs of the $\Phi$-equation the values of the 
graphs where the $\Box$-operators are distributed on the lower propagators are 
different. Explicit calculation shows that, at least to the order in $1/N$ 
which we are interested in, this is not the case. Indeed it turns out that 
these four graphs are straightforward to compute by the rules of conformal 
integration. The two non-planar graphs involve the same basic two loop integral
which we denote by $P(\mu)$. It is defined graphically in fig. 4 in the
coordinate space representation where one integrates over the location of the 
vertices. Thus, 
\begin{equation} 
\Pi_1 ~=~ \frac{a^2(1)}{a(2-\mu)} P(\mu) 
\end{equation} 
and 
\begin{equation} 
a(\mu-\alpha-\beta) \Sigma_1 ~=~ P(\mu) 
\end{equation} 
where we have included the compensating factor $a(\alpha+\beta)$ on the left
side of the value of the graph. This integral, $P(\mu)$, has not appeared in 
previous $O(1/N^3)$ calculations in other models and if we define 
\begin{equation} 
P(\mu) ~=~ \frac{a^2(\mu-1)}{(2\mu-3)} \Pi(\mu) 
\end{equation} 
then its $\epsilon$-expansion, where $d$ $=$ $4$ $-$ $2\epsilon$, is known to
very high orders, \cite{djbu62a,djbu62b}. In particular  
\begin{eqnarray} 
\Pi(\mu) &=& 6\zeta(3) ~+~ 9\zeta(4)\epsilon ~+~ 7\zeta(5)\epsilon^2 
{}~+~ \frac{5}{2}[\zeta(6) - 2\zeta^2(3)]\epsilon^3 
{}~-~ \frac{1}{8}[ 91\zeta(7) + 120\zeta(4)\zeta(3) ]\epsilon^4 
\nonumber \\ 
&& -~ \frac{1}{64}[ 5517\zeta(8) - 512\zeta(5)\zeta(3) - 768 U_{62} ]
\epsilon^5 ~+~ O(\epsilon^6) 
\end{eqnarray} 
The presence of the non-zeta transcendental at $O(\epsilon^5)$ encourages us
to believe there will be a $(3,4)$-torus knot in the wave function 
renormalization at $O(1/N^3)$ at some very high order in perturbation theory
when one disentangles the information encoded in the relation $\eta$ $=$ 
$\gamma_\Phi(g_c)/2$ using the $O(1/N^3)$ corrections to $g_c$ available in
the $\beta$-function critical exponent $\omega_2$ computed in \cite{pfjag}. For
$\Pi_2$ the integral in fact occurs in the $\beta$-function calculation of the 
Wess-Zumino model, \cite{pfjag}, as well as in the original work of 
\cite{vas12}. Therefore we merely quote the final result 
\begin{equation} 
\Pi_2 ~=~ \frac{a^2(1) a^2(2\mu-2)}{(\mu-2)^2}  
\left[ \frac{\hat{\Psi}^2(\mu)}{2} ~+~ \frac{\hat{\Phi}(\mu)}{2} ~+~ 
\frac{\hat{\Psi}(\mu)}{(\mu-2)} - 3\hat{\Theta}(\mu) \right] \eta_1^2  
\end{equation} 
where $\hat{\Theta}(\mu)$ $=$ $\psi^\prime(\mu-1)$ $-$ $\psi^\prime(1)$, 
$\hat{\Phi}(\mu)$ $=$ $\psi^\prime(2\mu-3)$ $-$ $\psi^\prime(3-\mu)$ $-$
$\psi^\prime(\mu-1)$ $+$ $\psi^\prime(1)$ and $\hat{\Psi}(\mu)$ $=$ 
$\psi(2\mu-3)$ $+$ $\psi(3-\mu)$ $-$ $\psi(\mu-1)$ $-$ $\psi(1)$. Therefore,
like $\Sigma_1$ and $\Pi_1$, $\Pi_2$ only has zeta transcendentals in its 
$\epsilon$-expansion. For $\Sigma_2$ elementary manipulations yield the 
three loop integral of fig. 5 which is illustrated in the coordinate space
representation in the case when $\delta$ $=$ $0$. It is evaluated by 
integration by parts using a temporary analytic regularization introduced by
shifting the exponents of some of the lines by an infinitesimal quantity,
$\delta$. This is necessary because the set of integrals which result are 
divergent though their sum is finite. The symmetric choice illustrated in 
fig. 5 is to minimize the number of subsequent graphs that need to be 
calculated as well as to ensure that these can still be computed by conformal
techniques. We find that  
\begin{equation} 
a(\mu-\alpha-\beta) \Sigma_2 ~=~ a(2\mu-2) \left[ \frac{5}{6} 
\hat{\Omega}(\mu) ~+~ 2\hat{\Theta}^2(\mu) \right] 
\end{equation} 
where $\hat{\Omega}(\mu)$ $=$ $\psi^{\prime\prime\prime}(\mu-1)$ $-$ 
$\psi^{\prime\prime \prime}(1)$. Hence we can now substitute for the values of 
the integrals in (\ref{eta3sd}) and expand out the leading term of 
(\ref{etait}) to $O(1/N^3)$ to deduce that in $d$-dimensions 
\begin{eqnarray} 
\eta_3 &=& \eta_1^3 \left[ \frac{(7\mu^2-26\mu+25)}{4(\mu-2)^2} 
\hat{\Psi}^2(\mu) ~+~ \frac{(3\mu^2-10\mu+9)}{4(\mu-2)^2}\hat{\Phi}(\mu) ~-~ 
(\mu-1)^2 \hat{\Theta}^2(\mu) \right. \nonumber \\ 
&& \left. ~~~~-~ \frac{5(\mu-1)^2}{12}\hat{\Omega}(\mu) 
{}~+~ \frac{(26\mu^4-174\mu^3+435\mu^2-479\mu+195)} 
{2(2\mu-3)(\mu-1)(\mu-2)^3}\hat{\Psi}(\mu) \right. \nonumber \\ 
&& \left. ~~~~-~ \frac{(7\mu^2-16\mu+10)}{4(\mu-2)^2} \hat{\Theta}(\mu) ~+~ 
\frac{(44\mu^4-257\mu^3+557\mu^2-531\mu+188)}{2(2\mu-3)^2(\mu-1)^2(\mu-2)^2} 
\right]  
\label{eta3} 
\end{eqnarray} 
In the final expression it turns out that the integral $P(\mu)$ has cancelled.
This is a direct result of the $D$-algebra, and therefore supersymmetry, which
modifies the exponents of each integral to be similar. A cancellation of some 
sort between $\Sigma_1$ and $\Pi_1$ was to have been expected, however, from 
knowledge of the $\epsilon$-expansion of the expression obtained for $\eta_3$ 
by ignoring the higher order contributions $\Sigma_i$ and $\Pi_i$. In other 
words that obtained by iterating (\ref{etait}) to $O(1/N^3)$. This 
approximation, known as Hartree Fock, in fact agrees with the known four loop 
perturbative result, (\ref{gamphi}), which implies that although $\Sigma_1$ and
$\Pi_1$ are non-zero, their $\zeta(3)$ contribution at three loops must at 
least cancel as well as $\zeta(3)$ and $\zeta(4)$ at four loops. (We note the 
Hartree Fock contribution to $\eta_4$ also agrees with the four loop 
perturbative result.) It could have been the case that $\Sigma_1$ and $\Pi_1$ 
took different values but had $\epsilon$-expansion agreement at low orders. 
Indeed this is the situation at four loops for $\Sigma_2$ and $\Pi_2$ whose low
order $\epsilon$-expansions are the same, correctly involving $\zeta(5)$ as the
first term but differing first at level $\zeta(7)$. They give consistency of 
the Hartree Fock $\eta_3$ to four loops but new non-trivial contributions will 
enter at five and higher loops. Therefore the explicit cancellation of the 
$P(\mu)$ function is novel and indicates that unlike its underlying bosonic 
$O(N)$ $\phi^4$ theory, there will be no non-zeta knots at all orders in the 
wave function renormalization at $O(1/N^3)$. Thus the presence of four 
dimensional supersymmetry would appear to reduce the number of complicated 
knots which will arise. This is in contrast to the case of the $O(N)$ 
supersymmetric $\sigma$ model in two dimensions where it is expected that its 
$\eta_3$ retains the integral yielding a $(3,4)$-torus knot, \cite{sig3,bgk}. 
One can understand this better by considering the simple properties of 
supersymmetric theories in differing dimensions. For instance, a four 
dimensional ${\cal N}$ $=$ $1$ supersymmetric scalar theory involves chiral 
superfields but in two dimensions it is theories with ${\cal N}$ $=$ $2$ 
supersymmetry which are chiral and not ${\cal N}$ $=$ $1$ theories such as the 
$O(N)$ supersymmetric $\sigma$ model. Therefore this would suggest that 
theories with ${\cal N}$ $=$ $2$ in two dimensions should not have non-zeta
knots at $O(1/N^3)$. 

Although (\ref{eta3}) clearly will agree with the $4$-loop perturbative result 
for $\gamma_\Phi(g)$ since the higher order graphs give non-zero contributions 
first at five loops, one question arises as to whether the correct signs and 
weighting factors have been included. However, the topologies of figs 2 and 3 
arise in the computation of the critical exponent which determines the 
$\beta$-function at $O(1/N^2)$, \cite{pfjag}. Therefore we have been careful to
include the corrections to the Schwinger Dyson equations with the same relevant
factors as in \cite{pfjag}, since these were essential to gain agreement with 
the $4$-loop perturbative $\beta$-function, (\ref{beta}). In light of these 
remarks we can now deduce several of the higher order coefficients of 
$\gamma_\Phi(g)$. If we denote the $O(1/N^3)$ part of $\gamma_\Phi(g)$ as 
\begin{equation} 
\gamma_\Phi(g) ~=~ g ~+~ \sum_{r=2}^\infty (c_r N^2 + d_r N + e_r)N^{r-3} g^r 
\end{equation} 
with $e_2$ $\equiv$ $0$ then it is easy to deduce that  
\begin{eqnarray} 
c_5 &=& \frac{1}{16}[ 3\zeta(4) - 2\zeta(3) - 1 ] \nonumber \\ 
d_5 &=& \frac{1}{24} [ 18\zeta(4) - 30\zeta(3) - 29 ] \\  
e_5 &=& \frac{1}{24} [ \, - \, 450\zeta(6) + 1416\zeta(5) - 171\zeta(4) 
+ 102\zeta(3) - 108\zeta^2(3) + 304 ] \nonumber  
\end{eqnarray} 
and 
\begin{eqnarray} 
c_6 &=& \frac{1}{32}[ 6\zeta(5) - 3\zeta(4) - 2\zeta(3) - 1 ] \nonumber \\ 
d_6 &=& \frac{1}{80}[ 80\zeta(5) - 162\zeta(4) + 248\zeta(3) - 81 ] \\ 
e_6 &=& \frac{1}{20} [ \, - \, 230\zeta(7) + 1140\zeta(6) - 906\zeta(5) 
- 36\zeta(4) - 5\zeta(3) - 108\zeta(3)\zeta(4) + 414\zeta^2(3) + 14] \nonumber 
\end{eqnarray} 
Therefore there are only two unknown coefficients to be determined in the 
polynomial of $N$ to complete the five loop result. Interestingly the first 
contribution to the perturbative coefficients from the higher order four loop
diagrams of figs 2 and 3 is the $\zeta(7)$ term at six loops. Although we have 
expressed our result in $d$-dimensions we cannot quote a value for $\eta_3$ in 
three dimensions since the expression diverges there. This is on a par with the 
critical exponent $\omega$ $=$ $-$ $\beta^\prime(g_c)$ which is also singular 
in three dimensions but at $O(1/N^2)$, \cite{pfjag}.  

In conclusion although we have ruled out non-zeta knots at $O(1/N^3)$ in the 
wave function renormalization of the $O(N)$ Wess-Zumino model it is not clear 
if this would be the case at all orders in perturbation theory or for the other 
renormalization group functions. For instance, one way of ascertaining whether
the $(3,4)$-torus knot survives at six loops in, say, the $\beta$-function
would be to analyse those Feynman diagrams with no subgraph divergences which
contributed in $\phi^4$ theory to its $\beta$-function, \cite{dkdb}, and 
determine their value in the context of the Wess-Zumino model in superspace. 
Indeed it may be the case that all of these topologies are excluded at the 
first stage by the chirality of the fields. Alternatively if that is not the 
case then it would be an intriguing exercise to see how the $D$-algebra 
modifies the sum of the contributions from each diagram and if there is an
overall cancellation.  
 
\vspace{0.25cm} 
{\bf Acknowledgements.} This work was carried out with support from PPARC
through an Advanced Fellowship, (JAG), and JNICT by a scholarship, (PF). We
also thank Prof D.R.T. Jones and Dr I. Jack for several useful discussions.

\newpage 
{\epsfxsize=12cm 
\epsfbox{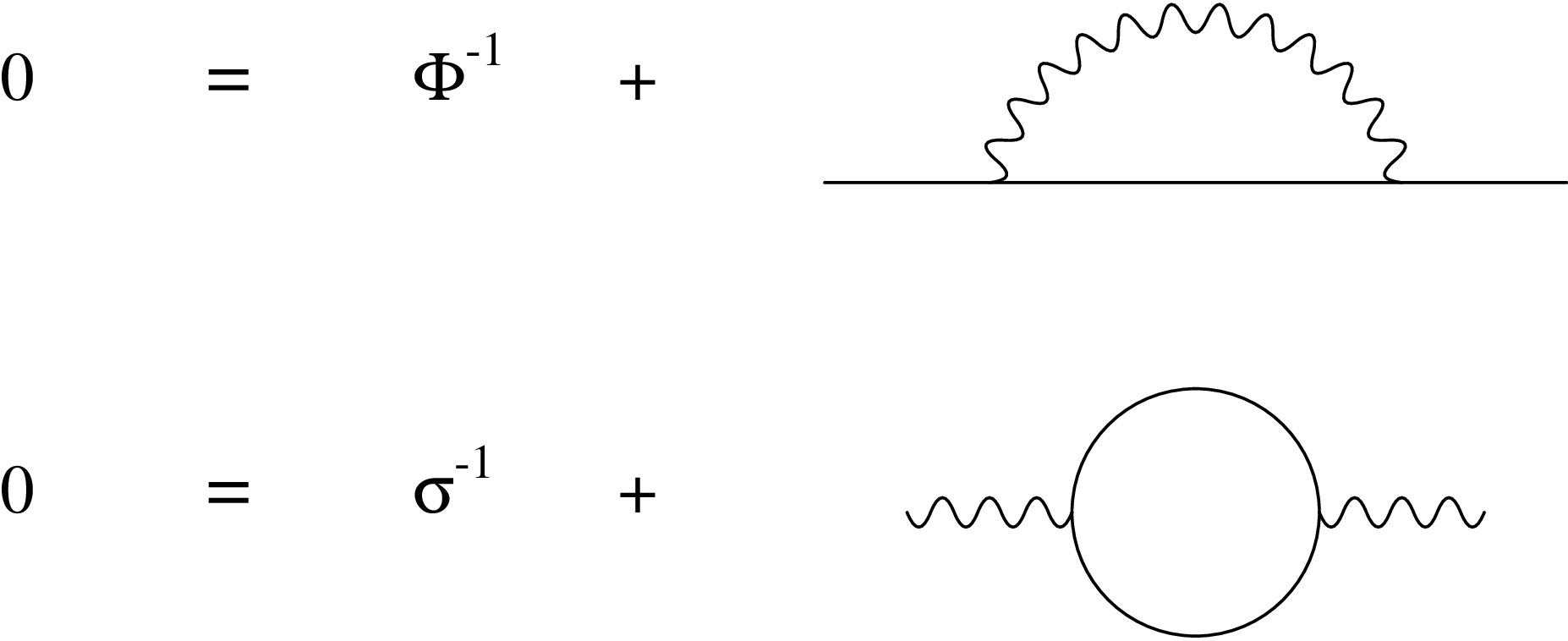}}  
\vspace{1cm} 
{\bf Fig. 1. Leading order Schwinger Dyson equations.} 

\vspace{3cm} 
{\epsfysize=3cm 
\epsfbox{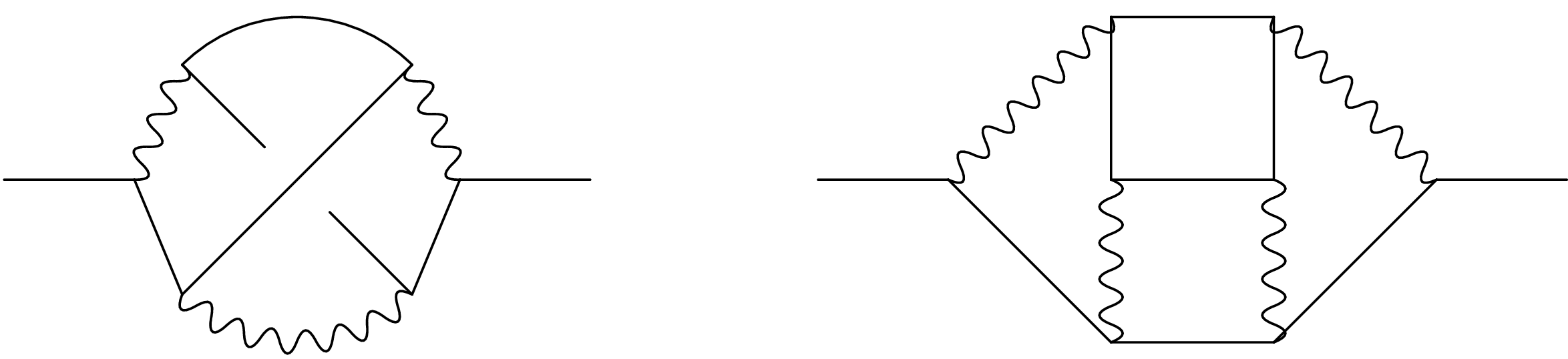}}  
\vspace{1cm} 
{\bf Fig. 2. Additional graphs contributing to the $\Phi$ Dyson equation at 
$O(1/N^3)$.} 

\vspace{3cm} 
{\epsfysize=3cm 
\epsfbox{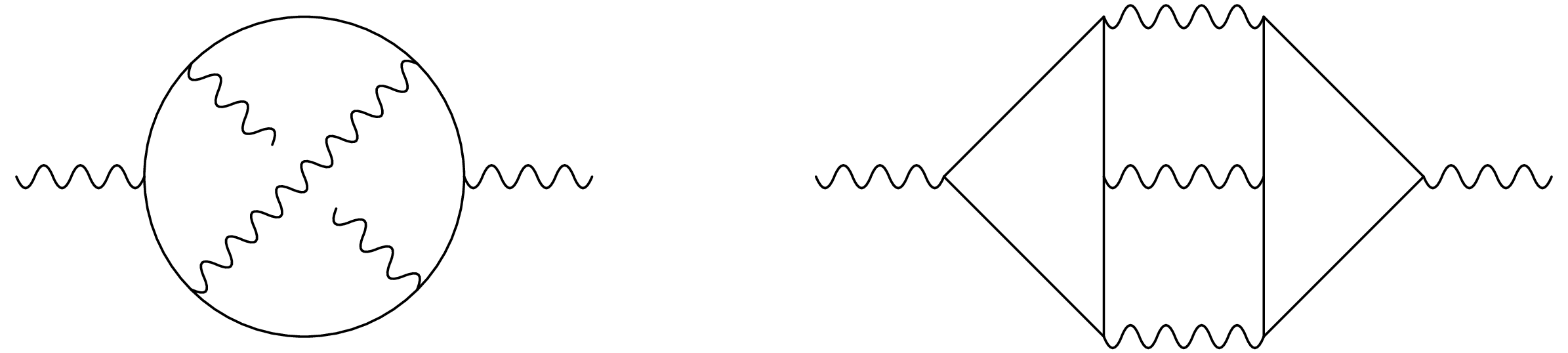}}  
\vspace{1cm} 
{\bf Fig. 3. Additional graphs contributing to the $\sigma$ Dyson equation at
$O(1/N^3)$.} 

\newpage 
{\epsfysize=3cm 
\epsfbox{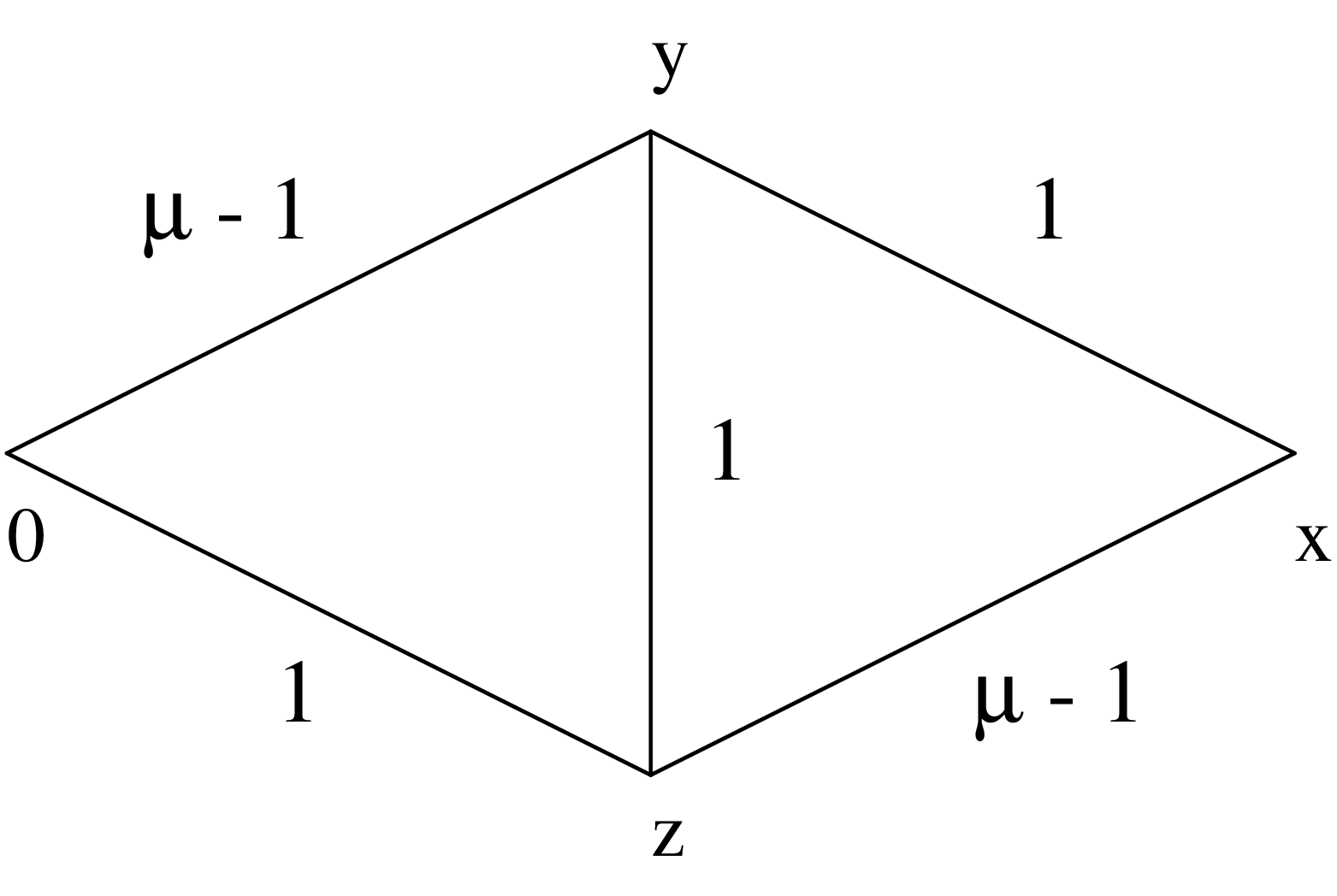}}  
\vspace{1cm} 
{\bf Fig. 4. Definition of the integral $P(\mu)$.} 

\vspace{3cm} 
{\epsfysize=4cm 
\epsfbox{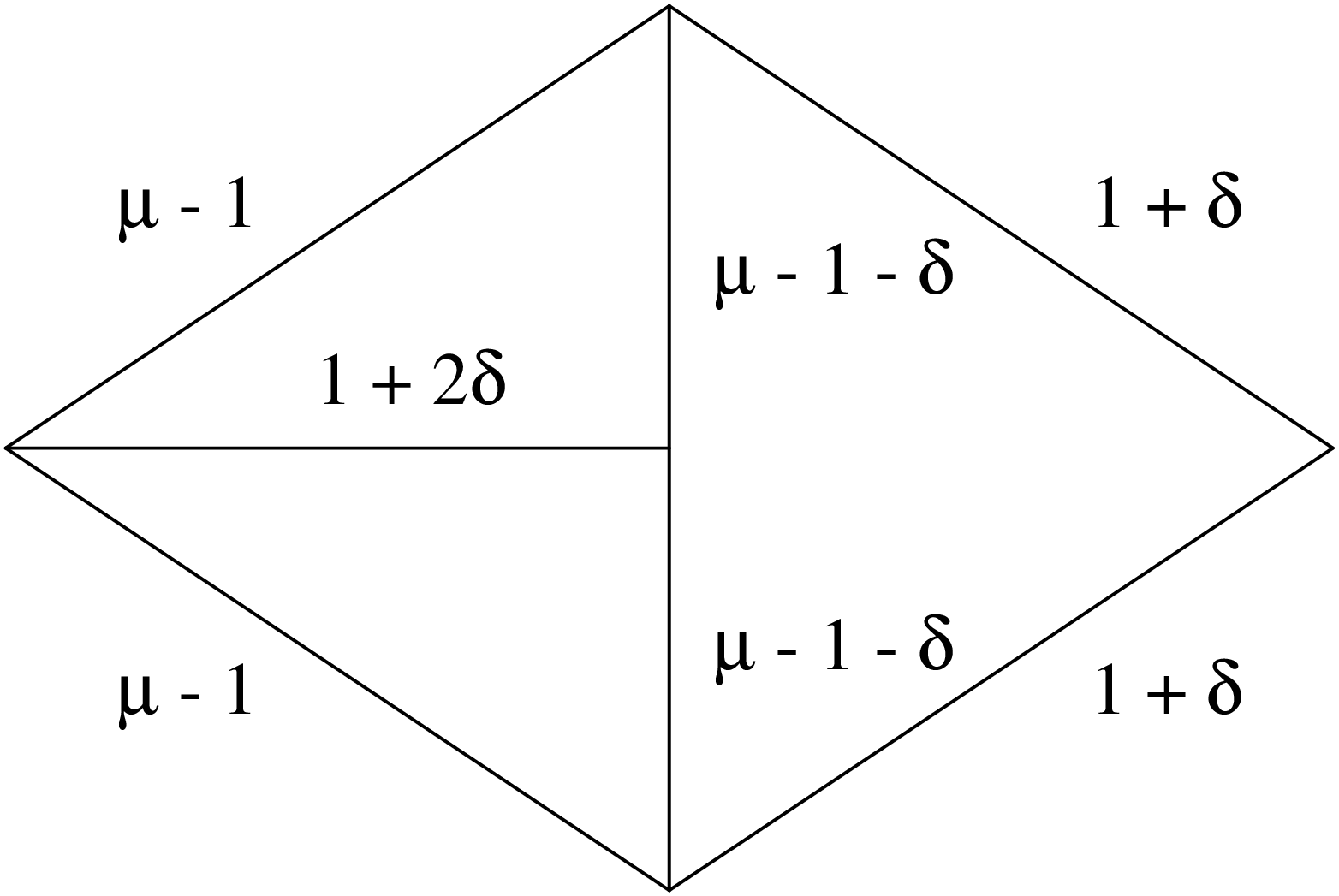}}  
\vspace{1cm} 
{\bf Fig. 5. Intermediate integral in the calculation of $\Sigma_2$.} 

\end{document}